\documentclass[reprint,aps,prl,twocolumn,superscriptaddress]{revtex4-2}

\usepackage{orcidlink}
\usepackage{graphicx}
\usepackage{amsfonts}
\usepackage{amsmath}
\usepackage{amssymb}
\usepackage{bbm}
\usepackage{blindtext}  
\usepackage{bm}
\usepackage{braket}
\usepackage{bbold}
\usepackage{color}
\usepackage{comment}
\usepackage{hyperref}
\usepackage{latexsym}
\usepackage{multirow}
\usepackage{subfigure}
\usepackage{soul}
\usepackage{tikz}
\usepackage{tabularx}
\usepackage{upgreek}
\usepackage{verbatim}
\usepackage{xcolor}

\hypersetup{
    colorlinks=true,
    linkcolor=blue,
    urlcolor=blue,
    filecolor=blue,
    citecolor=blue
    }

\makeatletter
\renewcommand{\p@subsection}{}
\renewcommand{\p@subsubsection}{}
\makeatother

\renewcommand{\H}{\hat{\mathcal{H}}}
\renewcommand{\a}{\hat{a}^{\phantom\dagger}}
\newcommand{\ad}{\hat{a}^\dagger}

\newcommand{\n}{\hat{n}}
\renewcommand{\S}{\hat{\mathbf{S}}}
\renewcommand{\P}{\hat{\mathcal{P}}}

\newcommand{\SuppMat}{
        \setcounter{table}{0}
        \renewcommand{\thetable}{S\arabic{table}}
        \setcounter{figure}{0}
        \renewcommand{\thefigure}{S\arabic{figure}}
        \setcounter{equation}{0}
        \renewcommand{\theequation}{S\arabic{equation}}
        \setcounter{page}{1}
    }

\begin{document}

% \title{Kinetic magnetism and stripe order in the doped AFM bosonic $t-J$ model}
\title{Kinetic magnetism and stripe order in the antiferromagnetic bosonic $t-J$ model}
%\thanks{A footnote to the article title}

\author{Timothy J. Harris\hspace{0.5mm}\orcidlink{0000-0001-9603-494X}}
\email{Tim.Harris@physik.uni-muenchen.de}
\affiliation{Department of Physics and Arnold Sommerfeld Center for Theoretical Physics (ASC), Ludwig-Maximilians-Universit\"at M\"unchen, Theresienstr. 37, M\"unchen D-80333, Germany}
\affiliation{Munich Center for Quantum Science and Technology (MCQST), Schellingstr. 4, M\"unchen D-80799, Germany}

\author{Ulrich Schollw\"{o}ck\hspace{0.5mm}\orcidlink{0000-0002-2538-1802}}
\affiliation{Department of Physics and Arnold Sommerfeld Center for Theoretical Physics (ASC), Ludwig-Maximilians-Universit\"at M\"unchen, Theresienstr. 37, M\"unchen D-80333, Germany}
\affiliation{Munich Center for Quantum Science and Technology (MCQST), Schellingstr. 4, M\"unchen D-80799, Germany}

\author{Annabelle Bohrdt\hspace{0.5mm}\orcidlink{0000-0002-3339-5200}}
\affiliation{Munich Center for Quantum Science and Technology (MCQST), Schellingstr. 4, M\"unchen D-80799, Germany}
\affiliation{University of Regensburg, Universit\"{a}tsstr. 31, Regensburg D-93053, Germany}

\author{Fabian Grusdt\hspace{0.5mm}\orcidlink{0000-0003-3531-8089}}
\email{Fabian.Grusdt@physik.uni-muenchen.de}
\affiliation{Department of Physics and Arnold Sommerfeld Center for Theoretical Physics (ASC), Ludwig-Maximilians-Universit\"at M\"unchen, Theresienstr. 37, M\"unchen D-80333, Germany}
\affiliation{Munich Center for Quantum Science and Technology (MCQST), Schellingstr. 4, M\"unchen D-80799, Germany}
\date{\today}

%%%%%%%%%%%%%%%%%%%%%%%%%%%%%%%%%%

\begin{abstract}
% Abstract blah blah... . 

% Unraveling the microscopic mechanisms governing the physics of doped quantum magnets is a central challenge in strongly correlated many-body physics, with implications for phenomena such as high-temperature superconductivity. Quantum simulation platforms, e.g., ultracold atoms in optical lattices or tweezer arrays, provide novel opportunities to disentangle the role of particle statistics in such strongly correlated phases. In this work we seek 

% Using large-scale density matrix renormalization group (DMRG) simulations, we map out the phase diagram of the AFM bosonic $t-J$ model on the 2D square lattice at finite doping. At low doping, bosonic holes form stripe patterns akin to those observed in high-$T_c$ cuprates. As doping increases, the system transitions from an AFM ground state to a ferromagnetic (FM) phase, driven by the motion of mobile bosonic charge carriers. At higher doping or large $t/J$, the system evolves into a fully-polarized SU(2) ferromagnet. Our findings shed new light on the role of particle statistics in strongly correlated quantum matter and offers an experimentally accessible platform to study spin-charge interplay. 

Unraveling the microscopic mechanisms governing the physics of doped quantum magnets is key to advancing our understanding of strongly correlated quantum matter. Quantum simulation platforms, e.g., ultracold atoms in optical lattices or tweezer arrays, provide a powerful tool to investigate the interplay between spin and charge motion in microscopic detail. Here, in a new twist, we disentangle the role of particle statistics from the physics of strong correlations by exploring the strong coupling limit of doped \emph{bosonic} quantum magnets, specifically the antiferromagnetic (AFM) bosonic $t-J$ model. Using large-scale density matrix renormalization group (DMRG) calculations, we map out the phase diagram on the 2D square lattice at finite doping. In the low-doping regime, bosonic holes form partially-filled stripes, akin to those observed in high-$T_c$ cuprates. As doping increases, a transition occurs to a partially-polarized ferromagnetic (FM) phase, driven by the motion of mobile bosonic charge carriers forming Nagaoka polarons. At high doping or large $t/J$, the system evolves into a fully-polarized ferromagnet. These findings shed new light on the role of particle statistics in strongly correlated many-body systems, revealing connections to stripe formation and the physics of kinetic (i.e., Nagaoka-type) ferromagnetism. Our results may be realized in state-of-the-art quantum simulation platforms with bosonic quantum gas microscopes and Rydberg atom tweezer arrays, paving the way for future experimental studies of doped bosonic quantum magnets.

\end{abstract}

%%%%%%%%%%%%%%%%%%%%%%%%%%%%%%%%%%

\maketitle

%%%%%%%%%%%%%%%%%%%%%%%%%%%%%%%%%%

% Developing a precise theoretical description of the fundamental mechanisms underpinning emergent many-body phenomena in strongly-correlated quantum systems (e.g. high-temperature superconductors) remains a one of the key outstanding challenges in the field of strongly correlated many-body physics 
% \cite{leeDopingMottInsulator2006}.

% \section{Introduction}\label{sec:introduction}

\textit{Introduction.}---Developing a precise theoretical descri- ption of the complex interplay between spin and charge degrees-of-freedom in doped Mott insulators is a central challenge in strongly correlated many-body physics 
\cite{leeDopingMottInsulator2006}. To date, the majority of the theoretical discussions have focussed on two paradigmatic Hamiltonians---namely the two-dimensional (2D) Fermi-Hubbard and related $t-J$ models---which are believed to capture the essential low-energy physics of high-$T_c$ cuprate superconductors \cite{emeryTheoryHighTcSuperconductivity1987,zhangEffectiveHamiltonianSuperconducting1988}. While these models have been pivotal in advancing our understanding of the rich variety of intertwined phases \cite{keimerQuantumMatterHightemperature2015}, e.g., through impressive recent progress in large-scale numerical simulations \cite{simonscollaborationonthemany-electronproblemSolutionsTwoDimensionalHubbard2015, simonscollaborationonthemany-electronproblemAbsenceSuperconductivityPure2020, xuCoexistenceSuperconductivityPartially2024}, it currently remains unclear to what extent the fermionic nature of the underlying charge carriers drives the emergence of these phases, as opposed to the competition between motional and magnetic degrees-of-freedom \cite{khatamiEffectParticleStatistics2012}. Addressing this distinction is crucial for uncovering the mechanisms underlying high-$T_c$ superconductivity and related phases, as it disentangles the role of particle statistics from the intrinsic physics of strong correlations.   

% Recent advances in analog quantum simulation platforms, e.g., ultracold atoms and molecules trapped in optical lattices or tweezer arrays, offer a unique platform for exploring this interplay in 

In this context, analog quantum simulation platforms, e.g., ultracold atoms trapped in optical lattices or tweezer arrays, provide a powerful approach to exploring the role of particle statistics in the formation of striped and superconducting phases \cite{bohrdtExplorationDopedQuantum2021, hirtheMagneticallyMediatedHole2023,bourgundFormationStripesMixeddimensional2023}. These systems provide well-isolated and fully-tunable realizations of minimal condensed matter models, allowing experiments to systematically probe the physics of strongly correlated ultracold quantum matter in microscopic detail \cite{grossQuantumSimulationsUltracold2017,browaeysManybodyPhysicsIndividually2020,kaufmanQuantumScienceOptical2021}. Key advantages of these platforms include precise control over particle statistics, lattice geometry, doping and interactions, as well as high-fidelity state preparation and measurement protocols facilitated by spin and charge resolved detection and addressing techniques \cite{weitenbergSinglespinAddressingAtomic2011, koepsellRobustBilayerCharge2020,bornetEnhancingManyBodyDipolar2024,hollandDemonstrationErasureConversion2024,picardSiteSelectivePreparationMultistate2024a,ruttleyEnhancedQuantumControl2024}. When combined with quantum gas microscopes \cite{grossQuantumGasMicroscopy2021,bohrdtExplorationDopedQuantum2021,christakisProbingSiteresolvedCorrelations2023},  these simulators enable the measurement of multi-point correlation functions that go beyond the capabilities of traditional solid-state experiments \cite{hilkerRevealingHiddenAntiferromagnetic2017,chiuStringPatternsDoped2019,koepsellMicroscopicEvolutionDoped2021,impertroLocalReadoutControl2024,schlomerLocalControlMixed2024}.

Building on this approach, here we study the antiferromagnetic (AFM) \textit{bosonic} $t-J$ model, which describes the interplay of magnetic and motional degrees-of-freedom in a system of mobile bosonic holes with local AFM interactions. This model offers a unique opportunity to examine the impact of bosonic statistics in a setting similar to that of fermionic models. Moreover, the AFM bosonic $t-J$ model can be realized in systems of ultracold atoms \cite{bohrdtMicroscopyBosonicCharge2024}, polar molecules \cite{gorshkovTunableSuperfluidityQuantum2011, carrollObservationGeneralizedTJ2024,homeierAntiferromagneticBosonicTJ2024}, and Rydberg tweezer arrays \cite{homeierAntiferromagneticBosonicTJ2024}, making it an experimentally accessible platform for testing theoretical predictions.

%At low densities (high doping) and weak interactions, the physics of bosonic dopants represents a radical departure from the principles of Fermi liquid theory. Indeed, as shown by Eisenberg and Lieb \cite{eisenbergPolarizationInteractingBosons2002}, spinful bosonic systems favor the formation of fully-polarized FM phases. 

Previous theoretical studies of the bosonic $t-J$ model have been restricted to lower-dimensional systems, cases with only \emph{partial} AFM couplings (e.g., $J_z>0, J_\perp \leqslant 0$), or in the high-temperature regime \cite{boninsegniPhaseSeparationMixtures2001,boninsegniPhaseSeparationStripes2002,smakovStripesTopologicalOrder2004,boninsegniPhaseDiagramAnisotropic2008,aokiMagneticOrderBoseEinstein2009,nakanoFinitetemperaturePhaseDiagram2011,nakanoFinitetemperaturePhaseDiagram2012,dickePhaseDiagramMixeddimensional2023}. However, the low-temperature (ground state) phase diagram at strong coupling ($t/J \gg 1$) with fully AFM spin-exchange interactions remains largely unexplored, particularly in the numerically challenging \emph{finite doping} regime. 

% At low densities (high doping) and weak interactions, the physics of bosonic dopants represents a radical departure from the principles of Fermi liquid theory. Indeed, as shown by Eisenberg and Lieb \cite{eisenbergPolarizationInteractingBosons2002}, spinful bosonic systems favor the formation of fully-polarized FM phases. 

In this work, we establish the $T=0$ phase diagram of the AFM bosonic $t-J$ model on the 2D square lattice at finite doping using large-scale density matrix renormalization group (DMRG) calculations \cite{schollwockDensitymatrixRenormalizationGroup2011, schollwockDensitymatrixRenormalizationGroup2005, whiteDensityMatrixFormulation1992,hubigSymmetryprotectedTensorNetworks2017,hubigSyTenToolkit} on finite-size cylinders \cite{stoudenmireStudyingTwoDimensionalSystems2012}. At low doping, i.e., close to the Heisenberg AFM---where the ground state is independent of the underlying statistics of the charge carriers---our numerical simulations unveil signatures of stripe order associated with incommensurate AFM correlations, reminiscent of the physics of doped fermionic Hubbard and $t-J$ models 
\cite{corbozStripesTwodimensionalTJ2011,zhengStripeOrderUnderdoped2017,qinHubbardModelComputational2022}, establishing a connection between our results and the broader physics of doped Mott insulators. In this regime, the ground state remains a total spin-singlet.

Beyond a critical doping value, $\delta>\delta^*_\mathrm{PP}$, the interplay between Heisenberg-type AFM superexchange interactions and itinerant Nagaoka-type ferromagnetism (FM), mediated by the motion of mobile bosonic holes, drives a transition to a partially-polarized (PP) FM phase, signified by a departure from a spin-singlet ground state. This transition is accompanied by the formation of so-called \emph{Nagaoka polarons}---i.e., regions of enhanced FM correlations embedded in an otherwise AFM spin background---resulting in a strong suppression of local AFM correlations as we dope away from half-filling. These Nagaoka polarons are a hallmark of the competition between spin and charge degrees-of-freedom in strongly correlated systems and directly connect our results to the physics of kinetic (i.e., Nagaoka-type) ferromagnetism \cite{nagaokaFerromagnetismNarrowAlmost1966} recently explored in quantum simulations of the triangular lattice Hubbard model using ultracold fermions \cite{prichardDirectlyImagingSpin2024,lebratObservationNagaokaPolarons2024}, arrays of quantum dots \cite{dehollainNagaokaFerromagnetismObserved2020} and moiré heterostructures \cite{tangSimulationHubbardModel2020, ciorciaroKineticMagnetismTriangular2023,taoObservationSpinPolarons2024}.

\begin{figure}[t!!]
\centering
\hspace*{0.2cm}
\includegraphics[width=0.74\linewidth]{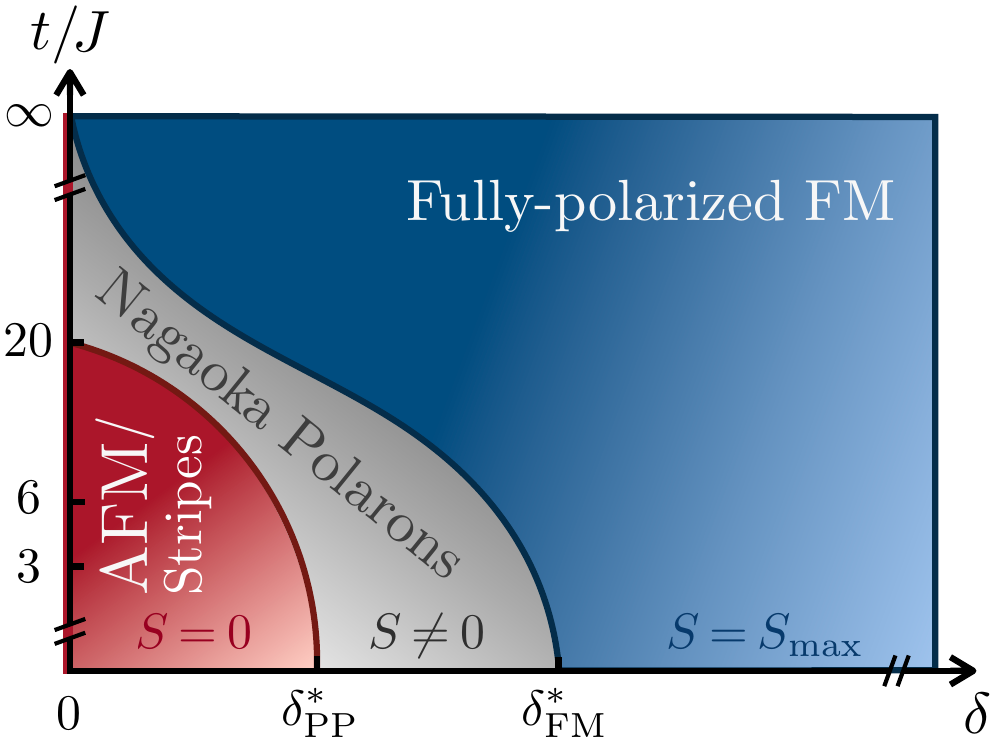}
\vspace{-2mm}
\caption{Schematic $T=0$ phase diagram of the AFM bosonic $t-J$ model indicating qualitatively different regions as a function of doping $\delta$ and interaction strength $t/J$. At half-filling (i.e., $\delta = 0$), the ground state is a long-range ordered Heisenberg AFM (red shaded region). At low doping, we also observe signatures of partially-filled stripes. As we increase doping, we identify a transition at $\delta=\delta^*_\mathrm{PP}$ to a partially-polarized Nagaoka polaron regime (grey) and then to a fully-polarized FM phase (blue) at $\delta=\delta^*_\mathrm{FM}$, characterized by total spin quantum numbers $S\neq 0$ and $S=S_\mathrm{max}$, respectively. Along the $y$-axis, for $t/J \gtrsim 20$ ($t/J\rightarrow \infty$) previous numerical studies \cite{whiteDensityMatrixRenormalization2001} have shown that a single dopant is sufficient to partially (or completely) polarize the system.}
\vspace{-4mm}
\label{Fig1}
\end{figure}

In the limit of large doping (i.e., for dilute systems)~or for large $t/J$, spatially overlapping Nagaoka polarons begin to percolate throughout the system, leading to the formation of a fully-polarized long-range SU(2) ferromagnetic ground state for $\delta > \delta^*_\mathrm{FM}$ \cite{lebratObservationNagaokaPolarons2024, samajdarPolaronicMechanismNagaoka2024}. Our results are summarized in Fig.~\ref{Fig1}.

\textit{The model.}---We consider a model of two-component (i.e., spin-$1/2$) hard-core bosons on the 2D square lattice:\\[-3mm]
\begin{equation}\label{bosonic_tJ_model}
    \begin{split}
        \H_{t-J} = -t&\sum_{\braket{\mathbf{i},\mathbf{j}}}\sum_\sigma\P_G\left(\ad_{\mathbf{i}\sigma}\a_{\mathbf{j}\sigma} + \mathrm{H.c.}\right)\P_G\\
        + J&\sum_{\braket{\mathbf{i},\mathbf{j}}}\left(\S_{\mathbf{i}}\cdot \S_{\mathbf{j}} + \frac{3}{4}\n_{\mathbf{i}}\n_{\mathbf{j}}\right),
    \end{split}
\end{equation}\\[-1.5mm]
\noindent where $\ad_{\mathbf{i}\sigma} (\a_{\mathbf{i}\sigma})$ are the creation (annihilation) operators for a boson on lattice site $\mathbf{i}=(i_x,i_y)$ in spin state $\sigma = \uparrow,\downarrow$; $\n_{\mathbf{i}} = \sum_{\sigma}\ad_{\mathbf{i}\sigma} \a_{\mathbf{i}\sigma}$ and $\S_{\mathbf{i}}$ are the local density and spin operators, respectively. Here $\braket{\mathbf{i},\mathbf{j}}$ denotes nearest-neighbor (NN) lattice sites, $t$ is the NN tunneling matrix element, $J$ is the isotropic Heisenberg superexchange coupling and $\P_G$ is the Gutzwiller projection operator which projects out all states with double occupancies. The bosonic $t-J$ model features both global $\mathrm{SU}(2)$ spin rotation symmetry and global $\mathrm{U}(1)$ particle number conservation symmetry, which we exploit in our numerical calculations \cite{supp_mat}. 

%%%%%%%%%%%%%%%%%%%%%%%%%%%%%%%%%%

%%%%%%%%%%%%%%%%%%%%%%%%%%%%%%%%%%

In the following, we fix the total particle number $N = N_\ell-N_h$, where $N_\ell=L_x\times L_y$ is the number of lattice sites and $N_h$ the total number of holes doped into the system (i.e., $\delta = N_h/N_\ell$). Moreover, we work in the regime with fully AFM spin-exchange interactions, $J>0$, where the model exhibits a sign problem \cite{dickePhaseDiagramMixeddimensional2023} and set $J=1$.

% Could instead define the hole doping level \delta = \sum_i\braket{\n^h_i}/N_\ell > 0

At half-filling ($\braket{\n_{\mathbf{i}}}=1$), the kinetic motion of holes is frozen out and the ground state of the model, Eq.~\eqref{bosonic_tJ_model}, is a long-range ordered Heisenberg AFM \cite{fazekasLectureNotesElectron1999}. In this regime, the ground state at strong coupling is identical for both fermionic and bosonic Hubbard systems with AFM interactions. Moreover, in the case of only a single dopant, it can be shown analytically that the ground state remains independent of the underlying particle statistics. However, in the \emph{finite doping} regime, i.e., where multiple holes are present, the distinction between fermionic and bosonic charge carriers becomes crucial. In this case, the system's behavior is governed by the competition between the kinetic motion of holes and AFM superexchange interactions, leading to distinct correlated phases for fermionic and bosonic dopants.

% \section{Results}\label{sec:results}

% \subsection{Ground state correlations at finite doping}\label{subsec:ground-state}

%%%%%%%%%%%%%%%%%%%%%%%%%%%%%%%%%%

\begin{figure*}[t!!]
\centering
\includegraphics[width=\textwidth]{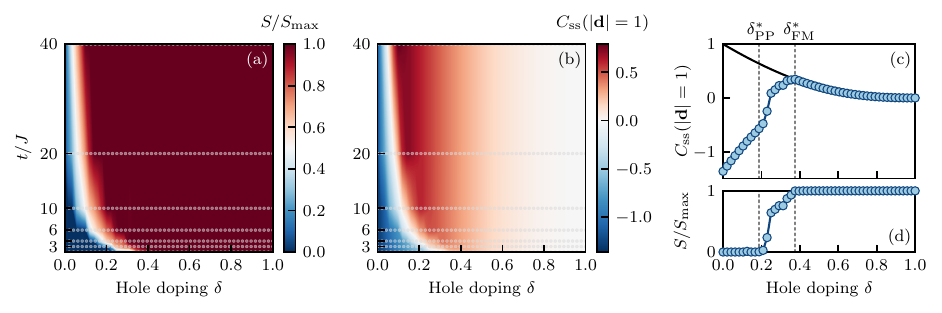}
\vspace{-0.7cm}
\caption{Ground state phase diagram of the AFM bosonic $t-J$ model. (a) We compute the total spin quantum number~$S/S_\mathrm{max}$ as a function of hole doping $\delta$ and $t/J$ obtained via DMRG simulations in the $S_z=0$ symmetry sector (data points denoted by light gray circles). (b) A qualitatively similar phase diagram is obtained by analyzing the NN spin--spin correlation function $C_\mathrm{ss}(|\mathbf{d}|=1)$, also as a function of hole doping and $t/J$. In (c),~(d) we show horizontal cuts through the data presented in~(a),~(b) for $t/J=3$, illustrating a clear transition between the AFM, partially-polarized and fully-polarized FM regimes. The positions of the critical doping values $\delta^*_\mathrm{PP}$ and $\delta^*_\mathrm{FM}$ are denoted by vertical gray dashed lines. In the fully-polarized FM phase, i.e., for $\delta > \delta^*_\mathrm{FM}\approx 0.38$, the NN correlations decay as $C_\mathrm{ss} \propto (1-|\delta|)^2$; the solid black line shows the correlations obtained from numerical simulations of a doped ferromagnet in the fully-polarized $S=S_\mathrm{max}$ symmetry sector. All DMRG calculations are performed on cylinders of size $L_x\times L_y = 16\times 6$.}
\label{Fig2}
\end{figure*}

%%%%%%%%%%%%%%%%%%%%%%%%%%%%%%%%%%

\textit{Ground state correlations at finite doping.}---We compute ground state properties of the doped AFM bosonic $t-J$ model, Eq.~\eqref{bosonic_tJ_model}, via applying the DMRG algorithm \cite{schollwockDensitymatrixRenormalizationGroup2011,schollwockDensitymatrixRenormalizationGroup2005,whiteDensityMatrixFormulation1992,hubigSyTenToolkit,hubigSymmetryprotectedTensorNetworks2017} on finite cylinders \cite{stoudenmireStudyingTwoDimensionalSystems2012} up to size $L_x\times L_y = 32\times 6$ with open (periodic) boundaries along the $x-$($y-$) direction(s). We explicitly perform calculations which implement both global $\mathrm{U}(1)_N \otimes \mathrm{SU(2)}_S$ and $\mathrm{U(1)}_N\otimes\mathrm{U(1)}_{S_z}$ symmetries to systematically verify the reliability of our results, and keep up to $\chi \sim 10,000$ SU(2) multiplets (i.e., equivalent to $\sim 30,000$ U(1) states) \cite{supp_mat}.   

% performed calculations which implement both global $\mathrm{U(1)}_N\otimes\mathrm{SU(2)}_S$ and $\mathrm{U(1)}_N\otimes\mathrm{U(1)}_{S_z}$ symmetries, in order to systematically verify the reliability of our numerical results.

% \subsubsection{Kinetic magnetism and Nagaoka polarons}

\textit{Kinetic magnetism and Nagaoka polarons.}---To characterize the magnetic nature of the ground state at finite doping, we evaluate the total spin squared via summing over two-point spin--spin correlation functions:\\[-4mm] 
\begin{equation}
    \braket{\hat{\mathbf{S}}^2} = \sum_{\mathbf{r}_1,{\mathbf{r}_2}=1}^{N_\ell}\braket{\S_{\mathbf{r}_1}\hspace{-1mm}\cdot\S_{\mathbf{r}_2}}.
\end{equation}\\[-3mm]
\noindent Computing the correlation functions numerically allows us to access the total spin $S$ of the ground state via the relation $\braket{\S^2} = S(S+1)$. A long-range SU(2) ferromagnet is characterized by maximum total spin quantum number $S_\mathrm{max}=(N_\ell-N_h)/2$. In Fig.~\ref{Fig2}(a), we plot the evolution of the total spin $S/S_\mathrm{max}$, normalized by its maximum value, as a function of doping for several values of $t/J$. We can clearly identify a regime close to half-filling ($\delta \approx 0$) where superexchange interactions dominate, resulting in a spin-singlet ground state with total spin $S=0$. 

As doping increases away from half-filling, the ground state transitions at $\delta = \delta^*_\mathrm{PP}$ to a partially-polarized (PP) phase, characterized by total spin $S\neq 0$. At even higher doping, specifically for  $\delta=\delta^*_\mathrm{FM} > \delta^*_\mathrm{PP}$, the system transitions into a fully-polarized FM phase where the total spin quantum number reaches its maximum value, $S=S_\mathrm{max}$. The fully-polarized FM phase remains stable even at very high doping, which distinguishes our findings from previous numerical studies of itinerant magnetism in geometrically frustrated fermionic systems \cite{moreraItinerantMagnetismMagnetic2024}.

% As doping increases beyond a critical value $\delta \gtrsim \delta^*_\mathrm{PP}$---which is dependent on the value of $t/J$---the ground state initially transitions to a partially-polarized, and then,~for $\delta>\delta^*_\mathrm{FM} > \delta^*_\mathrm{PP}$, to a fully-polarized FM, characterized by total spin quantum numbers $S\neq 0$ and $S=S_\mathrm{max}$, respectively. The fully-polarized FM phase is stable out to very large values of the doping $\delta$, which is a key difference between our results compared to previous numerical studies exploring itinerant magnetism in geometrically frustrated fermionic systems \cite{moreraItinerantMagnetismMagnetic2024}.

To connect with possible experiments, we compute~the connected spin--spin correlation function\\[-3mm]
\begin{equation}
    C_\mathrm{ss}(\mathbf{d}) = \frac{4}{\mathcal{N}_\mathrm{\mathbf{d}}}\sum_{\mathbf{r}_1-\mathbf{r}_2 = \mathbf{d}}\left(\braket{\S_{\mathbf{r}_1}\hspace{-1mm}\cdot\S_{\mathbf{r}_2}} - \braket{\S_{\mathbf{r}_2}}\hspace{-0.5mm}\cdot\hspace{-0.5mm}\braket{\S_{\mathbf{r}_2}}\right),
\end{equation}\\[-2mm]
\noindent where the normalization factor $\mathcal{N}_\mathbf{d}$ denotes the number~of lattice sites $\mathbf{r}_1,\mathbf{r}_2$ at a distance $\mathbf{d}$. In Fig.~\ref{Fig2}(b), we plot the NN spin--spin correlator $C_\mathrm{ss}(|\mathbf{d}|=1)$ as a function of doping $\delta$ for several ratios of $t/J$. At half-filling, superexchange interactions lead to a ground state characterized by strong NN AFM correlations (i.e., $C_\mathrm{ss}<0$). However, as we increase the doping level, this negative correlation is suppressed due to motion of mobile bosonic holes in the system (see Fig.~\ref{Fig2}(c)). For $\delta<\delta^*_{\mathrm{PP}}$, the ground state remains a total spin-singlet, as holes and spins interact in a way consistent with the formation of magnetic polarons or stripe-like order. Beyond the critical doping $\delta > \delta^*_{\mathrm{PP}}$, as we enter the partially-polarized FM phase, each hole binds to a finite bubble of non-zero magnetization, forming Nagaoka polarons. Indeed, the NN correlations eventually turn ferromagnetic ($C_\mathrm{ss}>0$) and saturate beyond a critical doping value $\delta \gtrsim \delta^*_\mathrm{FM}$, consistent with the picture of overlapping Nagaoka polarons driving a transition to a fully-polarized FM phase \cite{lebratObservationNagaokaPolarons2024}. For $\delta > \delta^*_\mathrm{FM}$, the NN correlations decay as $C_\mathrm{ss} \propto (1-|\delta|)^2$, as expected for a doped long-range SU(2) FM in this regime. Notably, the shift of $\delta^*_\mathrm{FM}$ to lower doping with increasing $t/J$ is consistent with the Nagaoka picture of a fully-polarized FM ground state emerging in the limit of infinitesimal positive doping as $t/J \rightarrow \infty$~\cite{nagaokaFerromagnetismNarrowAlmost1966}.

% To further explore the formation and spatial structure of the Nagaoka polarons driving the transition to an FM phase, we study the connected hole--spin--spin correlator:\\[-5mm] 

% \begin{equation}
%     C_\mathrm{hss}(\mathbf{r}_0;\mathbf{d}_1,\mathbf{d}_2) = \frac{4}{\mathcal{N}_\mathrm{hss}}\braket{\n^h_{\mathbf{r}_0}\S_{\mathbf{r}_0+\mathbf{d}_1}\hspace{-1mm}\cdot\S_{\mathbf{r}_0+\mathbf{d}_2}}_c,
% \end{equation}\\[-5mm]

% \noindent with normalization factor $\mathcal{N}_\mathrm{hss}$. The connected correlator $C_\mathrm{hss}$ quantifies the amount of spin correlations added by a hole to the normalized spin correlation background. 

% \subsubsection{Signatures of stripe order at low doping}

\begin{figure}[t!!]
\centering
\hspace*{0.3cm}
\includegraphics[width=\linewidth]{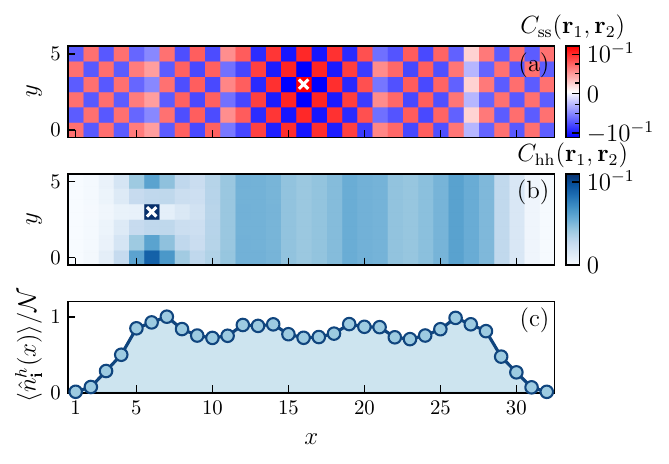}
\vspace{-0.8cm}
\caption{Numerical DMRG simulations signaling the existence of stripe order at low doping. (a) We plot the connected spin--spin correlation function $C_\mathrm{ss}(\mathbf{r}_1,\mathbf{r}_2)$ relative to the reference site $\mathbf{r}_1=(16, 3)$ for a $32\times 6$ cylinder at $t/J=3$ and $\delta=1/24$ in the $S=0$ sector. Correlations are color coded using a symmetric logarithmic scale, with linear interpolation between $-10^{-4}$ and $10^{-4}$. Domain walls in the AFM background are indicated by neighboring pairs of lattice sites with like-signed correlations along the $x$-axis. (b) Normalized hole--hole correlation function $C_\mathrm{hh}(\mathbf{r}_1,\mathbf{r}_2)$ with respect to the reference site $\mathbf{r}_1=(6,3)$. Within a stripe holes repel one another around the short (periodic) direction of the cylinder. (c) Mean hole density $\braket{\hat{n}^h_\mathbf{i}(x)} = \sum_{y=1}^{L_y} \braket{\n^h_\mathbf{i}(x,y)}/L_y$, normalized to unity, as a function of position along the long (open) $x$-direction of the cylinder. The density profile exhibits a periodic modulation, with hole rich regions corresponding to positions of AFM domain walls identified in (a). Reference sites are denoted by white crosses in (a) and (b).}
\vspace{-4mm}
\label{Fig3}
\end{figure}

\textit{Signatures of stripe order at low doping.}---In addition to signatures of kinetic magnetism and Nagaoka polarons observed in the magnetic correlations at intermediate to high doping, we also identify evidence for the emergence of partially-filled stripes in the low-doping regime where the ground state is a total spin-singlet~(i.e.,~$0 < \delta < \delta^*_\mathrm{PP}$). To investigate this, we compute the normalized hole--hole correlation function\\[-3mm]
\begin{equation}\label{eq:hole-hole correlator}
    C_\mathrm{hh}(\mathbf{r}_1,\mathbf{r}_2) =  \braket{\n^h_{\mathbf{r}_1}\n^h_{\mathbf{r}_2}}/\braket{\n^h_{\mathbf{r}_1}}.
\end{equation}\\[-3mm]
\noindent The correlator $C_\mathrm{hh}$ is suppressed if the presence of a hole at lattice site $\mathbf{r}_1$ makes it less likely to find a second hole at site $\mathbf{r}_2$, while $C_\mathrm{hh}$ is enhanced if it is more likely \cite{hirtheMagneticallyMediatedHole2023}. In Fig.~\ref{Fig3}, we plot $C_\mathrm{hh}$ together with the spin--spin correlations and hole density along the long (open) direction of the cylinder. In the ground state, we observe clear signatures of the formation of partially-filled stripes, indicated by (i) a periodic modulation of the hole density along the cylinder, and (ii) the appearance of AFM domain walls at positions of maximum hole density. 

The hole--hole correlations (see Fig.~\ref{Fig3}(b)) indicate that individual bosonic dopants located within a single stripe repel one another along the short (periodic)~direction of the cylinder, while there is no apparent spatial preference for two holes located within different stripes. This is in line with analogous charge correlations detected for fermionic stripes on finite cylinders \cite{blatzTwodopantOriginCompeting2024}, but stands in contrast to the emergence of tightly-bound hole pairs detected in previous numerical studies of paired phases in the 2D Fermi-Hubbard and $t-J$ models \cite{whiteHolePairStructures1997,blatzTwodopantOriginCompeting2024}.

% \tjh{Insert discussion about significance of stripes in the phase diagram of the bosonic $t-J$ model and comparison to fermionic systems etc.}

% \section{Discussion}\label{sec:discussion}

% \subsection{Experimental realization}

\textit{Experimental realization.}---Our study is motivated by recent advances in experimental platforms, e.g., bosonic quantum gas microscopes \cite{bohrdtMicroscopyBosonicCharge2024}, Rydberg atom tweezer arrays \cite{homeierAntiferromagneticBosonicTJ2024} and ultracold polar molecules \cite{carrollObservationGeneralizedTJ2024}. Here, we outline a scheme to adiabatically prepare strongly correlated many-body states of doped bosonic quantum antiferromagnets in ultracold atom experiments. This approach utilizes the unique advantages of ultracold atom quantum simulators, including their high degree of (local) control and tunability, combined with near perfect isolation from the surrounding environment. 

For concreteness, we consider a quantum gas microsc- ope of $^{87}$Rb atoms in a 2D square lattice geometry, where the effective spin-1/2 degrees-of-freedom are encoded in a pair of hyperfine states, e.g., $\ket{\downarrow} \equiv \ket{F=1,m_F=-1}$ and $\ket{\uparrow}\equiv\ket{F=2,m_F=2}$. Ultracold bosons in optical lattices are already an established platform for studying spin physics, offering insights into both equilibrium and out-of-equilibrium dynamics  \cite{trotzkyTimeResolvedObservationControl2008,simonQuantumSimulationAntiferromagnetic2011,fukuharaMicroscopicObservationMagnon2013,fukuharaQuantumDynamicsMobile2013,hildFarfromEquilibriumSpinTransport2014,jepsenSpinTransportTunable2020,jepsenTransverseSpinDynamics2021, sunRealizationBosonicAntiferromagnet2021,kimAdiabaticStatePreparation2024}. Crucially, $^{87}$Rb atoms allow for the generation of strong, locally-programmable staggered fields, which can be employed to adiabatically prepare low-entropy states without crossing a phase transition, providing an additional advantage in realizing and exploring doped bosonic quantum magnets \cite{bohrdtMicroscopyBosonicCharge2024, supp_mat}.

% ultracold bosonic atoms in optical lattices are already an established platofm in the quantum simulation of magnetism, in particular tunable spin-1/2 models have previously been ralized via X, Y, Z. 

%%%%%%%%%%%%%%%%%%%%%%%%%%%%%%%%%%

%%%%%%%%%%%%%%%%%%%%%%%%%%%%%%%%%%

Natively, ultracold bosons in optical lattices realize the spin-1/2 Bose-Hubbard model with (repulsive) on-site interactions $U_{\sigma\sigma'}$ between atoms and tunneling amplitude $t$. Working in sufficiently deep optical lattices allows us to reach the strong coupling limit ($U_{\sigma\sigma'} \gg t)$, where the low-energy physics (i.e., without double occupancy) is effectively described by the bosonic $t-J$ model, Eq.~\eqref{bosonic_tJ_model}, denoted here as $\H(t,J)$, with \emph{ferromagnetic} superexchange coupling $J = -4t^2/U < 0$ \cite{duanControllingSpinExchange2003}. However, by leveraging the bounded nature of the spectrum in the low-energy sector, it is possible to ``flip" the sign of the Hamiltonian in order to experimentally realize \emph{antiferromagnetic} coupling between bosons \cite{garcia-ripollImplementationSpinHamiltonians2004,sorensenAdiabaticPreparationManybody2010,braunNegativeAbsoluteTemperature2013}. This amounts to observing that the highest-energy eigenstate of the FM bosonic $t-J$ model, $\H(t,J)$, corresponds to the ground state of the AFM $t-J$ Hamiltonian $-\H(t,J) = \H(t,-J)$, where we have exploited the fact that on a bipartite lattice the sign of the tunneling matrix element $t$ can be exchanged via a local gauge transformation $\a_{\mathbf{j}\sigma}\rightarrow (-1)^\mathbf{j}\a_{\mathbf{j}\sigma}$, leaving~the overall physics invariant \cite{bohrdtMicroscopyBosonicCharge2024}.

In practice, implementing this scheme requires preparation of the highest excited eigenstate of the FM Hamiltonian $\H(t,J)$ with high fidelity. At half-filling, this can be achieved by locally projecting a strong staggered field of strength $h_z \gg |J|$ onto the lattice, for which a 2D N\'{e}el state corresponds to the highest excited eigenstate of the system. Starting from this initial configuration, adiabatically ramping down the staggered field, i.e., $h_z \rightarrow 0$, then prepares the highest excited state of $\H(t,J)$. To prepare ground states of doped bosonic quantum AFMs imposes further constraints on the state preparation protocol, additionally requiring that all holes in the system can be locally controlled \cite{bohrdtMicroscopyBosonicCharge2024}. In this way, striped ground states could also be directly prepared.

% \tjh{What other experimental details are important to mention here? More detailed discussions can be deferred to the Supplementary Material.}

%%%%%%%%%%%%%%%%%%%%%%%%%%%%%%%%%%

% \section{Summary and outlook}\label{sec:conclusions}

% The experimental implementation of $t-J$ models in tweezer arrays opens up a promising new route towards exploring strongly-correlated quantum matter. Especially the access to arbitrary two dimensional geometries, the broad tunability of parameters and the long-range tunnellings are new features that we can add to the toolbox of analog quantum simulation of doped quantum magnets. 

\textit{Outlook.}---In this letter, we have analyzed the delicate interplay between kinetic and magnetic exchange mechanisms in the phase diagram of the doped AFM bosonic $t-J$ model. We identified a transition from AFM to FM ordered ground states as a function of both doping and $t/J$, driven by the proliferation of Nagaoka polarons through the system. Additionally, we unveiled signatures of partially-filled stripe order in the low-doping regime, providing further evidence of a connection between our model and the strongly correlated phases of the fermionic Hubbard and $t-J$ models. Our results may be observed in state-of-the-art quantum simulation platforms, including quantum gas microscopes of ultracold bosonic atoms in optical lattices \cite{bohrdtMicroscopyBosonicCharge2024}, Rydberg atom tweezer arrays \cite{homeierAntiferromagneticBosonicTJ2024} and systems of ultracold polar molecules \cite{carrollObservationGeneralizedTJ2024}. 

% Our results provide a new perspective on strongly correlated systems by suggesting that an experimentally realizable Hamiltonian—--characterized by strong spin-charge correlations but without fermionic statistics--—can exhibit similar phenomenology to the fermionic models. 

% \tjh{I think we probably need 1--2 unifying sentences here. Perhaps mentioning that this work paves the way to disentangling the influence of particle statistics in our understanding of strongly correlated quantum matter?? Or similar...}

Future work may include extending our results to finite temperatures ($T>0$), thereby quantifying the effects of thermal fluctuations on the observed phase diagram of the doped AFM bosonic $t-J$ model, and~exploring possible connections between low-temperature striped phases and the emergence of a pseudogap in doped bosonic systems \cite{wietekStripesAntiferromagnetismPseudogap2021, simkovicOriginFatePseudogap2024,schlomerGeometricFractionalizedFermi2024}. From a practical perspective, it would~also be interesting to investigate the influence of tunable hole--magnon couplings in the system \cite{jepsenSpinTransportTunable2020,leeObservationSpinSqueezing2024} (e.g., how these may~modify spin--charge correlations in the finite doping regime), as well as the feasibility of utilizing tools including engineered dissipation \cite{krausPreparationEntangledStates2008,verstraeteQuantumComputationQuantumstate2009,daleyQuantumTrajectoriesOpen2014,harringtonEngineeredDissipationQuantum2022,miStableQuantumcorrelatedManybody2024}, entropy redistribution \cite{kantianDynamicalDisentanglingCooling2018,yangCoolingEntanglingUltracold2020} and adiabatic cooling \cite{lubaschAdiabaticPreparationHeisenberg2011,schachenmayerAdiabaticCoolingBosons2015,mirasolaCoolingFermionsOptical2018,venegas-gomezAdiabaticPreparationEntangled2020,sunRealizationBosonicAntiferromagnet2021,dimitrovaManybodySpinRotation2023, bohrdtMicroscopyBosonicCharge2024} to prepare strongly correlated many-body states of doped bosonic AFMs in ultracold atom experiments.

%%%%%%%%%%%%%%%%%%%%%%%%%%%%%%%%%%

\textit{Note added.}---During preparation of this manuscript, we became aware of a closely related work by Zhang \textit{et al.}~\cite{zhangQuantuminterferenceinducedPairingAntiferromagnetic2024}, in which they use DMRG to explore the~phase~diagram of the AFM bosonic $t-J$ model. In contrast to our study, the authors focus on finite cylinders of~width~$L_y=4,8$ using a combination of canonical~ensemble and grand-canonical ensemble simulations, respectively, for~which they report a pair-density wave with commensurate AFM order instead of the striped phase with incommensurate AFM order that we report in cylinders of width $L_y=6$. Additionally, we note a slight difference in model parameters, with local density-density interaction $+\frac{3}{4}\n_\mathbf{i}\n_\mathbf{j}$ used in this work, compared to $-\frac{1}{4}\n_\mathbf{i}\n_\mathbf{j}$ in Ref.~\cite{zhangQuantuminterferenceinducedPairingAntiferromagnetic2024}. 

%%%%%%%%%%%%%%%%%%%%%%%%%%%%%%%%%%

\begin{acknowledgments}
We wish to thank Tizian Blatz, Immanuel Bloch, Antoine Browaeys, Lukas Homeier, Hannah Lange, Sebastian Paeckel, Lode Pollet, Henning Schl\"{o}mer, David Wei and Johannes Zeiher for valuable discussions. This research was supported by the Deutsche
Forschungsgemeinschaft (DFG, German Research Foundation)
under Germany’s Excellence Strategy EXC-2111 Grant
No. 390814868 and the European Research Council
(ERC) under the European Union’s Horizon 2020 research
and innovation programme (Grant Agreement No. 948141),
ERC Starting Grant SimUcQuam. T.J.H acknowledges funding by the Munich Quantum Valley (MQV) doctoral fellowship program, which is supported by the Bavarian state government with funds from the Hightech Agenda Bayern Plus. Numerical simulations were performed on the Arnold Sommerfeld Center for Theoretical Physics High-Performance Computing cluster and the KCS cluster at the Leibniz Supercomputing Center (LRZ). 

% All DMRG calculations were performed using the \textsc{SyTen} toolkit, developed and maintained by C. Hubig, F. Lachenmaier, N.-O. Linden, T. Reinhard, L. Stenzel, A. Swoboda, M. Grunder, S. Mardazad, F. Pauw and S. Paeckel. Information is available at \href{\https://syten.eu/}{www.syten.eu}.
\end{acknowledgments}

% \tjh{Acknowledge developers/maintainers of SyTen toolkit?}

%%%%%%%%%%%%%%%%%%%%%%%%%%%%%%%%%%

% \appendix

% \section{}

%%%%%%%%%%%%%%%%%%%%%%%%%%%%%%%%%%

%% references
%\section*{References}
\bibliography{references.bib}
\widetext

%%%%%%%%%%%%%%%%%%%%%%%%%%%%%%%%%%
% supplementary material
%%%%%%%%%%%%%%%%%%%%%%%%%%%%%%%%%%

\newpage
\pagebreak
\appendix
%\onecolumngrid
\widetext

\newpage

\SuppMat

\begin{center}
\textbf{\large Supplementary Materials:\ Kinetic magnetism and stripe order\\ in the doped AFM bosonic $t-J$ model}
\end{center}

\vspace{3mm}

In this supplementary material, we derive the relevant low-energy effective Hamiltonian that we investigate, i.e., the bosonic $t-J$ model, as well as provide further discussion regarding possible experimental realizations of our results in ultracold atom quantum simulators. Additionally, we outline important technical details related to the numerical methods utilized in our work, and provide further simulation results to support our conclusions.

% In this supplementary material, we describe in detail the numerical methods utilized in our work and provide further analysis of the convergence of our results.

%%%%%%%%%%%%%%%%%%%%%%%%%%%%%%%%%%

\section{I. The bosonic $t-J$ model}

The bosonic $t-J$ model, Eq.~\eqref{bosonic_tJ_model}, may be derived in the strong coupling limit (i.e., $U_{\sigma\sigma^\prime} \gg t$) of the two-component (i.e., spin-$1/2$) Bose-Hubbard model, describing two species of bosons (e.g., encoded in two different hyperfine states of $^{87}$Rb atoms) confined to the lowest Bloch band of an optical lattice \cite{jakschColdBosonicAtoms1998, duanControllingSpinExchange2003}:\\[-2mm]

\begin{equation}\label{eq:spin-1/2_Bose_Hubbard_model}
    \H_{\mathrm{BH}} = \H_t + \H_U = -\sum_{\braket{\mathbf{i},\mathbf{j}}}\sum_\sigma t_\sigma\left(\ad_{\mathbf{i}\sigma}\a_{\mathbf{j}\sigma} + \mathrm{H.c.}\right) + \frac{1}{2}\sum_{\mathbf{i},\sigma} U_{\sigma\sigma}\n_{\mathbf{i}\sigma}(\n_{\mathbf{i}\sigma}-1) + U_{\uparrow\downarrow}\sum_{\mathbf{i}}\n_{\mathbf{i}\uparrow}\n_{\mathbf{i}\downarrow},
\end{equation}\\[-5mm]

\noindent where $\ad_{\mathbf{i}\sigma} (\a_{\mathbf{i}\sigma})$ are the creation (annihilation) operators for a boson on lattice site $\mathbf{i}$ in spin state $\sigma = \ \uparrow,\downarrow$; $\n_{\mathbf{i}\sigma} = \ad_{\mathbf{i}\sigma} \a_{\mathbf{i}\sigma}$ is the local density operator for spin-$\sigma$ bosons, $t_\sigma$ is the state-dependent NN tunneling matrix element and $U_{\sigma\sigma'}$ are the inter and intraspecies on-site interactions. Here $\braket{\mathbf{i},\mathbf{j}}$ denotes pairs of nearest-neighbor (NN) lattice sites, and we note that the bosonic operators satisfy canonical commutation relations:\\[-2mm]

\begin{equation}
    [\a_{\mathbf{i}\sigma},\a_{\mathbf{j}\sigma'}] = [\ad_{\mathbf{i}\sigma},\ad_{\mathbf{j}\sigma'}] = 0, \quad [\a_{\mathbf{i}\sigma},\ad_{\mathbf{j}\sigma'}] = \delta_{\mathbf{i}\mathbf{j}}\delta_{\sigma\sigma'},
\end{equation}\\[-7mm]

\noindent where $\delta_{\mathbf{i}\mathbf{j}}, \delta_{\sigma\sigma'}$ denote Kronecker delta functions. In the strong coupling limit (i.e., $U_{\sigma\sigma^\prime} \gg t$), performing a~Schrieffer-Wolff transformation \cite{auerbachInteractingElectronsQuantum1994,fazekasLectureNotesElectron1999} yields the following low-energy effective Hamiltonian, i.e., the bosonic $t-J$ model:\\[-2mm]

\begin{equation}\label{eq:bosonic_tJ_model_inc_3s_terms}
    \begin{split}
        \H_{t-J} &= \H_t + \H_J + \H_\mathrm{3-site}\\
        &= -\sum_{\braket{\mathbf{i},\mathbf{j}}}\sum_\sigma t_\sigma\P_G \left(\ad_{\mathbf{i}\sigma}\a_{\mathbf{j}\sigma} + \mathrm{H.c.}\right)\P_G + \sum_{\braket{\mathbf{i},\mathbf{j}}} \left(J_z\hat{S}^z_{\mathbf{i}}\hat{S}^z_{\mathbf{j}} + \frac{J_\perp}{2}\left(\hat{S}^+_{\mathbf{i}}\hat{S}^-_{\mathbf{j}} + \mathrm{H.c.}\right)+ 
        V\n_{\mathbf{i}}\n_{\mathbf{j}} - \frac{W}{2}\left(\hat{S}^z_{\mathbf{i}}\n_{\mathbf{j}} + \mathrm{H.c.}\right)\right)\\
        &\textcolor{white}{=} \hspace{+1mm}-\hspace{-1.2mm}\sum_{\braket{\mathbf{i},\mathbf{j},\mathbf{k}}}\hspace{-1.2mm}\sum_\sigma \left(\frac{t_\sigma^2}{U_{\uparrow\downarrow}}\ad_{\mathbf{i}\sigma}\n_{\mathbf{j}\bar{\sigma}}\a_{\mathbf{k}\sigma} + \frac{t_\uparrow t_\downarrow}{U_{\uparrow\downarrow}} \ad_{\mathbf{i}\bar{\sigma}}\hat{S}^\sigma_{\mathbf{j}}\a_{\mathbf{k}\sigma} + \frac{2t_\sigma^2}{U_{\sigma\sigma}}\ad_{\mathbf{i}\sigma}\n_{\mathbf{j}\sigma}\a_{\mathbf{k}\sigma} + \mathrm{H.c.}\right) + \mathcal{O}(t^3/U^2),\\
    \end{split}
\end{equation}\\[-8mm] 

% \noindent where $\P_G$ is the Gutzwiller projection operator which restricts charge motion to the low-energy Hilbert space (i.e., no double occupancy), $\braket{\mathbf{i},\mathbf{j},\mathbf{k}}$ denotes adjacent pairs of NN lattice sites $\mathbf{i},\mathbf{j}$ and $\mathbf{j},\mathbf{k}$ with $\mathbf{i}\neq \mathbf{k}$, $\S_\mathbf{i} = \sum_{\alpha\beta} \ad_{\mathbf{i}\alpha} \bm{\sigma}^{\phantom\dagger}_{\alpha\beta}\a_{\mathbf{i}\beta}/2$ are the local spin-1/2 operators with Pauli matrices $\bm{\sigma} = (\sigma_x,\sigma_y,\sigma_z)$ and $\alpha,\beta = \ \uparrow,\downarrow$. The local spin raising (lowering) operators are $\hat{S}^+_{\mathbf{i}}= \a_{\mathbf{i}\uparrow}\a_{\mathbf{i}\downarrow}$ ($\hat{S}^-_{\mathbf{i}} = \a_{\mathbf{i}\downarrow}\a_{\mathbf{i}\uparrow}$) with $\hat{S}^\sigma_{\mathbf{i}}\equiv\hat{S}^+_{\mathbf{i}} (\hat{S}^-_{\mathbf{i}})$ for $\sigma = \ \uparrow(\downarrow)$, and we introduce the Hamiltonian parameters:\\[-2mm] 

\noindent where $\P_G$ is the Gutzwiller projection operator which restricts charge motion to the low-energy Hilbert space (i.e., without double occupancy), $\braket{\mathbf{i},\mathbf{j},\mathbf{k}}$ denotes adjacent pairs of NN lattice sites $\mathbf{i},\mathbf{j}$ and $\mathbf{j},\mathbf{k}$ with $\mathbf{i}\neq \mathbf{k}$, $\overline{\sigma}$ flips the spin ($\overline{\uparrow}=\downarrow$ and vice versa), $\hat{S}^\sigma_{\mathbf{i}}\equiv\hat{S}^+_{\mathbf{i}} (\hat{S}^-_{\mathbf{i}})$ for $\sigma = \ \uparrow(\downarrow)$, and we have introduced the Hamiltonian parameters \cite{duanControllingSpinExchange2003,jepsenTransverseSpinDynamics2021}:\\[-0mm]

\begin{equation}
    \begin{split}
        J_z &= \frac{2(t_\uparrow^2+t_\downarrow^2)}{U_{\uparrow\downarrow}} - \frac{4t_\uparrow^2}{U_{\uparrow\uparrow}} - \frac{4t_\downarrow^2}{U_{\downarrow\downarrow}},\\
        J_\perp &= -\frac{4t_\uparrow t_\downarrow}{U_{\uparrow\downarrow}},\\
        V &= -\frac{t_\uparrow t_\downarrow}{U_{\uparrow\downarrow}}-\frac{t_\uparrow^2}{U_{\uparrow\uparrow}}-\frac{t_\downarrow^2}{U_{\downarrow\downarrow}},\\
        W &= \frac{4t_\uparrow^2}{U_{\uparrow\uparrow}}-\frac{4t_{\downarrow}^2}{U_{\downarrow\downarrow}}.\\[2mm]
    \end{split}
\end{equation}\\[-5mm]

\vfill

\noindent The Hamiltonian, Eq.~\eqref{eq:bosonic_tJ_model_inc_3s_terms}, describes hard-core bosonic particles tunneling on a lattice with NN magnetic interactions. The lower line describes second-order density-assisted and spin-flip-assisted NNN tunneling processes, respectively. In principle, both the sign and magnitude of the on-site interaction terms $U_{\sigma\sigma'}$ may be tuned via magnetic Feshbach resonance, as in Ref.~\cite{jepsenSpinTransportTunable2020}. However, for $^{87}$Rb atoms such a Feshbach resonance is not easily accessible, and we recover the approximate $\mathrm{SU}(2)$ symmetric limit, i.e., taking 
 $U_{\uparrow\uparrow} \approx U_{\downarrow\downarrow} \approx U_{\uparrow\downarrow}\equiv U$ and $t_\uparrow \approx t_\downarrow \equiv t$, and discarding three-site terms, which yields the bosonic $t-J$ Hamiltonian, Eq.~\eqref{bosonic_tJ_model}, with $J = -4t^2/U < 0$, $V = 3J/4 < 0 $ and $W=0$. 

%%%%%%%%%%%%%%%%%%%%%%%%%%%%%%%%%%

\newpage
\section{II. Experimental realization in a bosonic quantum gas microscope}

As outlined in the main text, the experimental setup we consider consists of a quantum gas microscope of ultracold bosonic $^{87}$Rb atoms confined to a single plane of a 3D optical lattice, thereby realizing a 2D square lattice geometry. For each atom, we encode an effective spin-1/2 degree-of-freedom in a pair of hyperfine states, e.g., $\ket{\downarrow} \equiv \ket{F=1,m_F=-1}$ and $\ket{\uparrow}\equiv\ket{F=2,m_F=2}$. This setup naturally realizes the two-component Bose-Hubbard model, Eq.~\eqref{eq:spin-1/2_Bose_Hubbard_model}, and in the strong coupling limit (i.e., $U_{\sigma\sigma'} \gg t)$, the low-energy physics is well described by the bosonic $t-J$ model, Eq.~\eqref{bosonic_tJ_model}, with \emph{ferromagnetic} superexchange coupling $J = -4t^2/U < 0$ \cite{duanControllingSpinExchange2003}. 

% We note, that a highly-tunable generalization of the AFM bosonic $t-J$ model, Eq.~\eqref{bosonic_tJ_model}, may also be realized by encoding the three-dimensional local Hilbert space, $\mathcal{H} \in \{\ket{h},\ket{\downarrow},\ket{\uparrow}\}$, where $\ket{h}$ denotes the hole state (equally the absense of a spin), into three internal atomic or molecular states as first proposed in Ref.~\cite{homeierAntiferromagneticBosonicTJ2024}. 
We propose to utilize metastable negative absolute-temperature states to engineer AFM couplings between bosons, following proposals in Refs.~\cite{garcia-ripollImplementationSpinHamiltonians2004,sorensenAdiabaticPreparationManybody2010,braunNegativeAbsoluteTemperature2013}. Specifically, we aim to prepare the highest-excited eigenstate of the FM bosonic $t-J$ model (see Sec.~II.~A for details), which on a bipartite lattice, e.g., the 2D square lattice, corresponds to the ground state of the AFM $t-J$ Hamiltonian. To see this, we note that on a bipartite lattice with sublattices $\mathcal{A}$ and $\mathcal{B}$, the sign of the tunneling matrix element $t$ can be exchanged via the local gauge transformation\\[-2mm]
\begin{equation}
    \a_{\mathbf{j}\sigma} \rightarrow (-1)^\mathbf{j}\a_{\mathbf{j}\sigma} = \begin{cases}
        +\a_{\mathbf{j}\sigma}, \quad \mathbf{j}\in\mathcal{A},\\
        -\a_{\mathbf{j}\sigma}, \quad \mathbf{j}\in\mathcal{B},
    \end{cases}
\end{equation}

\noindent and likewise for $\ad_{\mathbf{j}\sigma}$. This transformation leaves the magnetic interaction term $\H_J$ invariant, while mapping $\H_t \rightarrow -\H_t$. Therefore, when turning around the spectrum, it is only necessary to consider changing the sign of the spin-interaction in the FM Hamiltonian, $\H_J \rightarrow -\H_J$. Overall, we obtain that the high-energy physics of the FM bosonic $t-J$ model is described by $\H_{t-J}' = -\H_{t-J} = \H_{t} - \H_{J}$, which directly yields the AFM bosonic $t-J$ model introduced in Eq.~\eqref{bosonic_tJ_model}. 

\subsection{A. Adiabatic state preparation scheme}\label{sec:adiabatic_state_prep}
We propose an adiabatic ramp protocol to prepare strongly correlated many-body states of doped bosonic quantum magnets. In this case, these states correspond to the highest-excited eigenstates of the FM bosonic $t-J$ Hamiltonian. The full state preparation protocol is outlined as follows:

\begin{enumerate}
    \item Prepare a unit-filled spin-polarized Mott insulator of $^{87}$Rb atoms, i.e., $\ket{\Psi_\mathrm{Mott}} = \prod_\mathbf{i}\ket{\downarrow}_\mathbf{i}$, in a 2D square lattice.
    
    \item Deterministically prepare the desired charge state (i.e., total $N$ sector) via site-resolved addressing techniques.
    
    \item We then apply strong local light-shifts on only a single sublattice, e.g., the $\mathcal{A}$ sublattice, realizing the staggered field Hamiltonian $\H_z(\tau) = h_z(\tau)\sum_{\mathbf{j}}(-1)^{\mathbf{j}}\hat{S}^z_{\mathbf{j}}$. The total effective Hamiltonian is then given by:\\[-2mm]
    \begin{equation}\label{eq:tJ_model_w_staggered_field}
        \H(\tau) = \H_{t-J} + \H_z(\tau) = \H_{t-J} + h_z(\tau)\sum_{\mathbf{j}}(-1)^{\mathbf{j}}\hat{S}^z_{\mathbf{j}},
    \end{equation}\\[-2mm]
    \noindent where $h_z(\tau)$ is the time-dependent field strength. The staggered field in Eq.~\eqref{eq:tJ_model_w_staggered_field} may be realized (up to a global $\hat{S}_z$ term) via locally programmable ``anti-magic'' differential light shifts available for $^{87}$Rb \cite{bohrdtMicroscopyBosonicCharge2024}.
    
    \item Applying a global microwave $\pi$-pulse on the shifted $\ket{\downarrow}\leftrightarrow\ket{\uparrow}$ transition then resonantly transfers the remaining atoms on the $\mathcal{A}$ sublattice from $\ket{\downarrow}\rightarrow \ket{\uparrow}$, preparing a product state in the desired total $\hat{S}_z$ sector which is the highest-excited eigenstate of the effective Hamiltonian, Eq.~\eqref{eq:tJ_model_w_staggered_field}, for $h_z(\tau)> t \gg J$. For example, at half-filling, this protocol prepares the N\'{e}el state $\ket{\Psi_{\text{N\'{e}el}}} = \prod_{\mathbf{i}\in \mathcal{A}}\prod_{\mathbf{j}\in \mathcal{B}}\ket{\uparrow_\mathbf{i}\downarrow_\mathbf{j}}$ \footnote{The other N\'{e}el state, $\ket{\Psi_\text{N\'{e}el}} = \prod_{\mathbf{i}\in \mathcal{A}}\prod_{\mathbf{j}\in \mathcal{B}}\ket{\downarrow_\mathbf{i}\uparrow_\mathbf{j}}$, with sublattices $\mathcal{A}$ and $\mathcal{B}$ reversed, corresponds to the ground state of the effective Hamiltonian $\H(\tau)$.}. 
    
    \item Finally, we adiabatically ramp down the staggered field, $h_z(\tau)\rightarrow 0$, to prepare the ground state $\ket{\Psi_0}$ of $\H_{t-J}^{\mathrm{AFM}}$. During this process, dopants are pinned by \emph{anti-trapping potentials} in order to maximize the kinetic energy of the mobile bosonic holes. One example of a possible ramp profile, is given by an exponential ramp of the form $h_z(\tau) = h_0 e^{-\tau/T}$, parametrized by the total ramp time $T$. In the limit $T\rightarrow \infty$ the ramp is fully adibatic.
    \vfill
\end{enumerate}

Of course, the overall success of any (quasi-)adiabatic ramp protocol depends sensitively on the details of the low-energy eigenstates of the system. In particular, we note that the total ramp time $T$ should scale as $T \propto 1/\Delta^2$---with $\Delta$ the energy gap to the first excited state---in order to remain in the highest instantaneous eigenstate of the system.  In this case, in the idealized SU(2)-invariant limit, there is a gap closing exactly at $h_z=0$, which will result in a finite density of excitations in the final state for a finite preparation time. However, for sufficiently slow ramps, we expect this density to be small. A detailed investigation of realistic finite-time ramp protocols---which may be optimized via quantum control techniques \cite{xieBayesianLearningOptimal2022, blatzBayesianOptimizationRobust2024}---and the impact of experimental imperfections, including disorder, decoherence, state preparation and measurement (SPAM) errors, and thermal fluctuations, remains the subject of future work.

%%%%%%%%%%%%%%%%%%%%%%%%%%%%%%%%%%
% \newpage
\section{III. Numerical simulations using DMRG}

All ground state calculations in this work were performed using the DMRG algorithm \cite{schollwockDensitymatrixRenormalizationGroup2011,schollwockDensitymatrixRenormalizationGroup2005,whiteDensityMatrixFormulation1992} in its matrix product state (MPS) formulation as implemented via the \textsc{SyTen} toolkit \cite{hubigSyTenToolkit,hubigSymmetryprotectedTensorNetworks2017} on finite cylinders up to size $L_x \times L_y = 32 \times 6$ with open (periodic) boundaries along the $x-$($y-$) direction(s) \cite{stoudenmireStudyingTwoDimensionalSystems2012}. In particular, we utilize a combination of strictly single-site (DMRG3S) \cite{hubigStrictlySinglesiteDMRG2015} and two-site (2DMRG) update schemes---scaling as $\mathcal{O}(\chi^3 d w)$ and $\mathcal{O}(\chi^3 d^2 w)$ respectively, where $\chi$ is the MPS bond dimension,  $d$ the dimension of the local Hilbert space and $w$ is the matrix product operator (MPO) bond dimension---in order to ensure optimal convergence. 

% Convergence figure
%%%%%%%%%%%%%%%%%%%%%%%%%%%%%%%%%%

\begin{figure*}[t!!]
\centering
\includegraphics[width=0.6\textwidth]{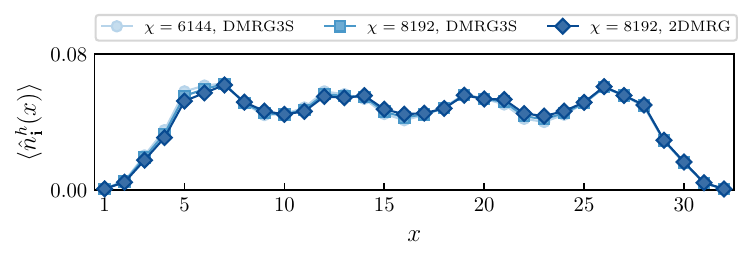}
\vspace{-4mm}
\caption{We show the mean hole density profile $\braket{n^h_{\mathbf{i}}(x)}=\sum_{y=1}^{L_y}\braket{n^h_{\mathbf{i}}(x,y)}/L_y$ along the long (open) direction of the cylinder for the last three stages of our DMRG sweeping procedure, corresponding to the results shown in Fig.~\ref{Fig3} of the main text. The lattice parameters are $L_x\times L_y = 32\times 6$, $t/J=3$ and $\delta =1/24$ and the simulation is performed in the $S=0$ symmetry sector using global $\mathrm{U(1)}_N\otimes\mathrm{SU(2)}_S$ symmetries. The maximum local MPS bond dimensions are given by $\chi = \{6144,8192,8192\}$, and we employ either strictly single-site (DMRG3S) or two-site (2DMRG) update schemes. The truncation error for the final stage (i.e., $\chi=8192$, 2DMRG) is $\mathcal{T}(\epsilon) = 2.86 \times 10^{-6}$ with left-right imbalance $\mathcal{I} < 1\%$.}
\vspace{-1.5mm}
\label{FigS1}
\end{figure*}

%%%%%%%%%%%%%%%%%%%%%%%%%%%%%%%%%%
\vspace{-1mm}
\subsection{A. Implementation of non-abelian symmetries}

The performance of the DMRG algorithm may be greatly enhanced by exploiting any available symmetries of the Hamiltonian under consideration. In addition to abelian symmetries that are available in standard open-source tensor network libraries, e.g., global U(1) particle number $N = \sum_{\mathbf{i}}\braket{\n_\mathbf{i}}$ and magnetization $S_z = \sum_{\mathbf{i}}\braket{\hat{S}^z_\mathbf{i}}$ conservation \cite{singhTensorNetworkStates2011}, the \textsc{SyTen} toolkit also allows for flexible implementation of non-abelian symmetries \cite{singhTensorNetworkStates2012,weichselbaumNonabelianSymmetriesTensor2012}, including global SU(2) spin-rotation symmetry corresponding to conservation of the total spin quantum number $S$. 

When utilizing non-abelian symmetries, the local MPS bond dimension $\chi_i$ required to faithfully represent a given quantum state is reduced to an effective bond dimension $\chi^*_i$ which counts multiplets only, i.e., groups of states each belonging to a given \emph{irreducible representation} of the underlying non-abelian symmetry. The resulting reduction in the local bond dimension from $\chi_i\rightarrow \chi_i^*$ is given by a factor of $\mathcal{R} \sim 3^{r}$, where $r$ is the \emph{rank} of the underlying symmetry group, e.g., $\mathcal{R} \sim 3^{N-1}$ for SU($N$) which has rank $r=N-1$ \cite{weichselbaumNonabelianSymmetriesTensor2012}. For more information about utilizing non-abelian symmetries in the context of tensor network calculations, we refer the interested reader to Refs.~\cite{singhTensorNetworkStates2012,weichselbaumNonabelianSymmetriesTensor2012}.

From a numerical point of view, the reduction in the effective local bond dimension $\chi^*_i$ afforded by the use of SU(2) spin-rotation symmetry in our calculations is incredibly valuable. In practice, for the Fermi/Bose-Hubbard and $t-J$ type Hamiltonians considered in this work, we find that this ratio is well aligned with the expected value $\mathcal{R} \approx 3$ when working in the $S=0$ symmetry sector (i.e., total spin-singlet). Since the computational cost of the DMRG algorithm scales as $\mathcal{O}(\chi^3)$, we therefore expect a computational speed-up of $\mathcal{O}(3^{3}) \sim 30$ for a simulation of comparable accuracy. 

There are some caveats to the practical benefits afforded by this approach, namely that when the total spin $S$ of the ground state is changing as a function of a tunable system parameter (e.g., hole doping $\delta$), then in order to correctly identify the ground state symmetry sector, (parallel) DMRG calculations must be performed in each of the accessible total $S$ sectors, resulting in an additional computational overhead. Alternatively, the total spin quantum number can be determined from calculations implementing global $\mathrm{U(1)}_N\otimes\mathrm{U(1)}_{S_z}$ symmetries, as shown in Fig.~\ref{Fig2} of the main text. In this work, we have explicitly performed calculations which implement both global $\mathrm{U(1)}_N\otimes\mathrm{SU(2)}_S$ and $\mathrm{U(1)}_N\otimes\mathrm{U(1)}_{S_z}$ symmetries, in order to systematically verify the reliability of our numerical results.

% Equally, for a fixed computational time, exploiting SU(2) symmetry allows us to obtain far more accurate results than if we simply implemented U(1) magnetization conservation. 

%%%%%%%%%%%%%%%%%%%%%%%%%%%%%%%%%%

\begin{figure*}[t!!]
\centering
\includegraphics[width=\textwidth]{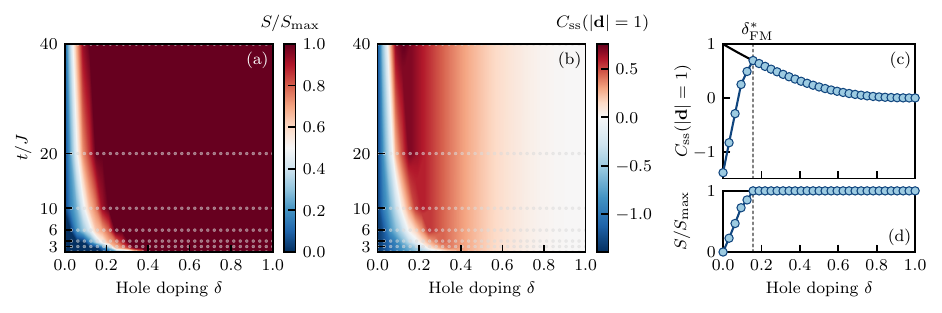}
\caption{Additional numerical results, as in Fig.~\ref{Fig2} from the main text, obtained via DMRG calculations in the $S_z=0$ symmetry sector performed on cylinders of size $L_x\times L_y = 16\times 4$. Heatmaps showing (a) the total spin quantum number $S/S_\mathrm{max}$ and (b) the NN spin--spin correlation function $C_\mathrm{ss}(|\mathbf{d}|=1)$ as a function of hole doping $\delta$ and $t/J$, respectively (data points are denoted by light gray circles). In (c),~(d) we show horizontal cuts through the data in~(a),~(b) for $t/J=20$. For $t/J=20$, $\delta^*_\mathrm{PP}\approx 0$ and the system is immediately partially polarized upon doping, as indicated by the departure from a spin-singlet ground state for $\delta\neq 0$. In the fully-polarized FM phase, i.e., $\delta > \delta^*_\mathrm{FM}\approx 0.16$, the NN correlations decay as $C_\mathrm{ss} \propto (1-|\delta|)^2$, consistent with the correlations obtained from numerical simulations of a doped ferromagnet in the $S=S_\mathrm{max}$ symmetry sector (solid black line).}
\label{FigS2}
\end{figure*}

%\tjh{The linear suppression of AFM correlations for $\delta < \delta^*_\mathrm{FM} \approx 0.18$ is consistent with the formation of Nagaoka polarons that begin to proliferate through the system}

%%%%%%%%%%%%%%%%%%%%%%%%%%%%%%%%%%
\vspace{-1mm}
\subsection{B. Analysis of convergence behavior}

We initialize our simulations using a randomly generated state in the desired symmetry sector, i.e., $(N,S_z)$ or $(N,S)$, before commencing our DMRG sweeps. We systematically increase the maximum bond dimension of our calculations until convergence is reached. For the ground state calculations shown in the main text, we utilized bond dimensions up to $\chi = 10,240$ (equivalent to $\chi \sim 30,000$ U(1) symmetric states for calculations using $\mathrm{SU(2)}$ spin-rotation symmetry) and obtain typical  truncation errors $\mathcal{T}(\epsilon)\sim \mathcal{O}(10^{-7})$ and $\mathcal{T}(\epsilon)\sim \mathcal{O}(10^{-6})$ for $L_y=4$ and $L_y=6$ width cylinders, respectively. To further ensure optimal convergence, we monitor local one- and two-body observables, including the local hole density, spin expectation values, hole--hole and spin--spin correlation functions. 

% \tjh{In particular, we find that $\braket{\hat{S}^z_{\mathbf{i}}}\sim\mathcal{O}(10^{-7})$ and $4\braket{\hat{S}^z_{\mathbf{i}}\hat{S}^z_{\mathbf{j}}} - \braket{\hat{S}^+_{\mathbf{i}}\hat{S}^-_{\mathbf{j}}+\mathrm{H.c.}}\sim\mathcal{O}(10^{-5})$, which are both expected to vanish identically due to the global $\mathrm{SU(2)}$-spin symmetry of the model, see Eq.~\eqref{bosonic_tJ_model}.}

In Fig.~\ref{FigS1}, we show the mean hole density $\braket{n^h_{\mathbf{i}}(x)}=\sum_{y=1}^{L_y}\braket{n^h_{\mathbf{i}}(x,y)}/L_y$ corresponding to the last three DMRG stages used to obtain the numerical results presented in Fig.~\ref{Fig3} of the main text. The hole density serves as a sensitive probe for numerical convergence, as slight real-space perturbations to the density profile are typically associated with very small energy cost. Moreover, the sweeping nature of the DMRG algorithm itself means that it is generally difficult to obtain a symmetric profile. We observe that in Fig.~\ref{FigS1}, the hole density is slightly asymmetric for $\chi= 6144$, but symmetric and almost indistinguishable for the final two stages with $\chi = 8192$, indicating that the final state is likely well converged. To quantify the asymmetry, we compute the imbalance $\mathcal{I} = (N_h^L - N_h^R)/N_h$, defined as the particle number difference between the left ($L$) and right ($R$) sides of the system, where $N_h^\sigma = \sum_{\mathbf{i}\in \sigma}\braket{\hat{n}^h_{\mathbf{i}}}$ for $\sigma = L,R$. We find that in the final DMRG stage, the hole density profile is indeed well converged, with left-right asymmetry $\mathcal{I} < 1\%$.

% \subsection{C.\quad Sign problem in quantum Monte Carlo simulations}

% A short discussion on why the bosonic $t-J$ model, Eq.~\eqref{bosonic_tJ_model}, exhibits a sign problem in quantum Monte Carlo~(QMC) calculations. Include a figure illustrating an exchange cycle leading to a sign problem as in Ref.~\cite{dickePhaseDiagramMixeddimensional2023}.

\subsection{C. Additional numerical results}

In Fig.~\ref{FigS2}, we show additional numerical results for the total spin quantum number and local spin--spin correlations obtained on cylinders of size $L_x\times L_y = 16\times 4$. We observe a qualitatively similar phase diagram for these calculations, as in Fig.~\ref{Fig2} of the main text, obtained for width-6 cylinders of size $L_x\times L_y = 16\times 6$. This indicates that the phase diagram we obtain is not strongly influenced by finite-size effects.

% \tjh{TO DO: Insert short discussion of qualitative agreement between results for $L_x\times L_y = 16\times 4$ and $16\times 6$ cylinders, i.e. refer to Fig.~\ref{FigS1}}.

\end{document}